\documentclass{rjparticle}
\usepackage{amssymb,latexsym,amsmath}
\usepackage{graphicx}

\def\rootfig{./}

\newcommand{\pt}{{\cal PT}}

\begin{document}

\title{
Nonlinear $\mathcal{PT}$-Symmetric models Bearing Exact Solutions}

\author[1]{H. Xu}
\affil[1]{Department of Mathematics and Statistics, University of Massachusetts,
Amherst MA 01003-4515, USA}

\author[1]{P.G.\ Kevrekidis }

\author[1]{Q. Zhou}

\author[2]{D.J. Frantzeskakis}
\affil[2]{Department of Physics, University of Athens, Panepistimiopolis, Zografos, GR-15784 Athens, Greece}

\author[2]{V. Achilleos}

\author[3]{R.\ Carretero-Gonz\'{a}lez}
\affil[3]{Nonlinear Dynamical Systems Group,
Computational Science Research Center, and
Department of Mathematics and Statistics,
San Diego State University, San Diego,
CA 92182-7720, USA}

\maketitle
\begin{abstract}
We study the nonlinear Schr{\"o}dinger equation with a $\pt$-symmetric potential.
Using a hydrodynamic formulation and
connecting the phase gradient to the field amplitude,
allows for a reduction of the model to a
Duffing or a generalized Duffing equation. This way, we can obtain exact soliton solutions
existing in the presence of suitable $\pt$-symmetric potentials, and study their stability and dynamics.
We report
interesting new features, including oscillatory instabilities of
solitons  and
(nonlinear) $\pt$-symmetry breaking
transitions, for focusing and defocusing nonlinearities.
\end{abstract}

\section{Introduction}

It is well known that in
quantum mechanics
the energy spectrum, as well as the spectrum of
operators associated with observables,
is real. This led to the commonly adopted assumption
that the Hamiltonian $\hat{H}$ must be Hermitian,
which guarantees a real spectrum.
However,
non-hermitian Hamiltonians respecting
$\mathcal{P}\mathcal{T}$-symmetry,
can also present entirely real spectra (within suitable parametric ranges) \cite{Bender,Bender2}.
The parity operator $\mathcal{P}$ corresponds to spatial reflections, $\hat{p}\to-\hat{p}$ and $\hat{x}\to-\hat{x}$, while
the time reversal operator $\mathcal{T}$ corresponds to
$\hat{p}\to-\hat{p}$, $\hat{x}\to\hat{x}$ and $i\to-i$.
%


While the subject of
$\mathcal{P}\mathcal{T}$-symmetric systems was initially
studied in the context of quantum mechanics, relevant
experimental realizations
emerged in other fields, most notably in optics.
In particular, such systems were proposed
\cite{Muga,ziad1,ziad2,Ramezani,Kuleshov} and subsequently experimentally implemented
\cite{dncnat,salamo} in systems of
optical waveguides. Another physical
context where such systems have been experimentally ``engineered''
is that of electronic circuits \cite{Schindler,tsampikos_review}. Recently, experiments on $\mathcal{P}\mathcal{T}$-symmetric media have also appeared in the context of whispering gallery modes\cite{Peng}.
In parallel to these significant
experimental developments, there has been an ever expanding
volume of theoretical works, considering both continuum and
discrete models (see, e.g., the recent review \cite{dumitru} and references therein).
Given the spatial limitations of
this contribution, we do not summarize these works here, but
rather focus on efforts relevant
to the considerations herein.

Already from
the early works studying the interplay between
$\pt$-symmetry and nonlinearity~\cite{ziad1,ziad2}, it became apparent
that exact solutions could be analytically obtained
for suitably ``carved'' potentials
---such as the Scarff-II potential \cite{ahmed}.
The idea of using Rosen-Morse
type $\pt$-symmetric potentials, and obtaining analytical solutions
thereof, was recently generalized to multiple potential
parameters and higher-dimensional cases in Ref.~\cite{Midya}%
\footnote{However,
in the latter work, the stability results reported are somewhat less
transparent,
partly due to the lack of dependence on the phase of the field in the
stability equations (2.8) therein, as well as due to the absence of
neutral or invariance associated modes in figures, such as Fig.~3(c)
therein.}.
More recently,
the relevant approach was revisited and generalized~\cite{Salerno},
using a decomposition into amplitude and phase,
for a $\pt$-symmetric term also in the nonlinear part
of the equation;
the emphasis was on producing an effective equation
for the density, with either a parabolic confining potential
or to construct effective equations with elliptic function
potentials, which
support exact analytical solutions.
Notice that the considerations in the latter case, were driven by the form of the
real part of the potential.

Here, we consider a system with a $\mathcal{P}\mathcal{T}$-symmetric Hamiltonian, namely
a one-dimensional (1D)
nonlinear Schr{\"o}dinger (NLS) equation with a complex potential:
%
\begin{align}
i u_t=-\frac{1}{2}u_{xx}+\sigma|u|^2u+[V(x)+iW(x)]u,
\label{pteqn1}
\end{align}
where $u$ is a complex field, subscripts denote partial derivatives,
$\sigma=+1$ ($\sigma=-1$) corresponds to a defocusing (focusing) nonlinearity, while
$V(x)$ and $W(x)$ correspond,
respectively,
to the real and imaginary (i.e., gain/loss) parts of the external
potential. Note that the system possesses
a $\pt$-symmetric Hamiltonian if $V(x)$
and $W(x)$ are, respectively, even and odd functions of $x$.
Our aim is to return to the ideas of Refs.~\cite{Midya,Salerno},
%
but from a different perspective. On the one hand, regarding
the existence problem, we consider cases with $V(x)=0$,
connecting the problem with the Duffing
equation or generalizations thereof (involving also higher exponents).
On the other hand, we also
place some emphasis on the stability analysis of the obtained
solutions, both for $\sigma=\pm 1$. Particularly, we
present some interesting features including the (unusual) emergence of
oscillatory instabilities for single bright solitons, or the
existence of a nonlinear $\pt$-phase transition between the homogeneous
and the dark soliton state for defocusing cubic-quintic nonlinearities.
For all the obtained solutions, upon
exploring their existence and stability, numerical simulations are
developed to appreciate their dynamical evolution.

Our presentation is structured as follows. In section II, we
focus on analytical considerations of the existence problem, providing
a systematic approach for generalized Duffing functional forms.
In section III, we explore some case examples of focusing and
defocusing cubic and cubic-quintic cases. Finally, in section IV,
we summarize our findings and present some future challenges.

\section{Theoretical Setup}


We seek
stationary solutions to Eq.~(\ref{pteqn1}) in the form
$u=\rho(x)\exp[i\phi(x)-i\mu t]$, where the amplitude $\rho(x)$ and phase $\phi(x)$ are real-valued
functions.
The resulting hydrodynamic-type equations for $\rho(x)$ and
$\phi$ read:
\begin{align}
&
\rho_{xx}-\rho\phi_x^2+2\mu\rho-2\sigma\rho^3-2V(x)\rho=0,
\label{s1} \\
&2W(x)\rho^2-(\rho^2\phi_x)_x=0.
\label{s2}
\end{align}
Our aim is to solve an ``inverse problem'',
i.e., to find $\mathcal{PT}$-symmetric potentials $W(x)$, for different choices of $\phi_x$.
This is done by prescribing the connection of $\phi_x$ with $\rho$, solving Eq.~(\ref{s1})
with respect to $\rho$, and determining $W(x)$ via Eq.~(\ref{s2}).
Here, we focus on the specific case of $\phi_x =\epsilon \rho^k$, where $k$ is an integer,
and $V(x)=0$; then, Eqs.~(\ref{s1})-(\ref{s2}) lead to
the system:
\begin{align}
&
\rho_{xx}-2\mu\rho-\epsilon^2\rho^{2k+1}-2\sigma\rho^3=0,
\quad
W(x)=
(1+k/2)\epsilon\rho^{k-1}\rho_x.
\label{pteqn5}
\end{align}
When $k$ is even, $W(x)$ is always an odd function. However, if $k$ is odd, we
need $\rho$ to be an even function to make $W(x)$ odd. Generally, the problem is reduced
to the solution of the ordinary differential equation (ODE)
\begin{align}
\rho_{xx}=a\rho+b\rho^m+c\rho^n,
\label{ode}
\end{align}
where coefficients and exponents depend on the choice of $k$.
While, in principle, our prescription can be carried out for any $k$, we have found it
progressively more difficult to identify exact solutions for larger
$k$, hence only cases with $k \leq 2$ are considered in what follows.

%


We start with the simplest case of $k=0$.
Then, the gain/loss profile reads
$W(x)
=\frac{\epsilon}{2}(\ln(\rho^2))_x$ and
Eq.~(\ref{ode})
becomes a simple Duffing model with
$\alpha=\mu-\epsilon/2$, $b=2\sigma$ and $c=0$.
The Duffing equation can be explicitly solved and its solutions
may have the form of solitons. In particular, for $\sigma=+1$, the
dark soliton solution reads
%
$\rho =\sqrt{\mu-{\epsilon^2}/{2}}\,\tanh(\sqrt{\mu-{\epsilon^2}x/{2}})$.
However, since $\rho(0)=0$, the obtained form for $W(x)$ diverges at $x=0$ and it is, thus, unphysical.
On the other hand,
for $\sigma=-1$, the
bright soliton
has the form $\rho =\sqrt{\epsilon^2-2\mu}\,\, {\rm sech}(\sqrt{\epsilon^2-2\mu} x)$.
In this case, $W(x)$ has a $\tanh$ profile, which describes
finite gain/loss
at $x\rightarrow\pm\infty$,
which again does not appear to be physical, or likely to lead to robust solutions.

Next, we consider the case of $k=1$ leading to
$W(x)=(3/2)\epsilon\rho_x$ and a Duffing equation for $\rho$
with $a=\mu$ and $b=2\sigma+\epsilon^2$.
This case, for $\sigma=+1$, yields:
%
\begin{align}
\rho(x)=\sqrt{\frac{\mu}{1+\epsilon^2/2}} \tanh(\sqrt{\mu}x), \quad
W(x)=\frac{3\mu}{2}\sqrt{\frac{\epsilon^2}{1+\epsilon^2/2}}{\rm sech}^2(\sqrt{\mu}x).
\label{pteqn3}
\end{align}
Since $W(x)$ is even, this case does not correspond to a $\pt$-symmetric system and will not be considered further.
%
On the other hand, for $\sigma=-1$, we obtain the results:
%
\begin{align}
\rho(x)=\sqrt{\frac{-2\mu}{1-\epsilon^2/2}} {\rm sech}(X),
\quad
W(x)=3\mu \sqrt{\frac{\epsilon^2}{1-\epsilon^2/2}}
{\rm sech}(X)
\tanh(X),
\label{pteqn4}
\end{align}
where $X=\sqrt{-2\mu}x$. This result corresponds
to a $\pt$-symmetric case, which was considered
both in linear \cite{ahmed} and nonlinear settings
(see, e.g., Refs.~\cite{ziad1,ziad2}).

Finally, we consider the $k=2$ case corresponding to
the cubic-quintic form
\begin{align}
\rho_{xx}=a\rho+b\rho^3+c\rho^5,
\label{cq}
\end{align}
with $a=-2\mu$, $b=2\sigma$ and $c=\epsilon^2$.
%
As a result, for
any solution of Eq.~(\ref{cq}) having a definite parity (including bright or dark solitons),
the ensuing $W(x)$ will be odd and, hence, of relevance
to our study of $\pt$ symmetric systems. We explore some special
cases where analytical
solutions can be obtained for this model below.

Analytical solutions of Eq.~(\ref{cq}) can be found as follows.
We introduce in this equation the transformation $\rho(x)=U(x)[g_1U^2(x)+g_2]^{-1/2}$, and
derive an ODE for $U(x)$. Then, requiring that $U(x)$ satisfies the Duffing equation:
\begin{align}
U_{xx} =
l_1U+l_3U^3,
\label{duf}
\end{align}
and its integrated counterpart $U_x^2=l_0+l_1U^2+\frac{l_3}{2}U^4$, we reduce the ODE for $U(x)$ to a
5th-order algebraic equation for $U$. The latter is satisfied if all coefficients are zero; this leads to
the following algebraic conditions:
\begin{align}
&g_1=\frac{b(l_1-a)}{2(l_1-a)^2+3l_0l_3-4l_1(l_1-a)}, ~~
g_2=\frac{3l_0 b}{2(l_1-a)^2+3l_0l_3-4l_1(l_1-a)}, \label{g12}
\\
&b^2\left[\frac{9 l_0 l_3}{l_1-a}-(4l_1+2a)\right]+2c\left[2(l_1-a)+\frac{3l_0l_3}{l_1-a}-4l_1\right]^2=0.
\label{pteqn8}
\end{align}
Then, solutions of Eq.~(\ref{cq}) can be constructed by solutions of the Duffing Eq.~(\ref{duf}).
%
%

Let us first consider the dark soliton solution of Eq.~(\ref{duf}),
$U=\tanh(\alpha x)$. In this case,
$l_0=\alpha^2$, $l_1=-2\alpha^2$ and $l_3=2\alpha^2$; then,
$\alpha$ can be found via
Eq.~(\ref{pteqn8}) and $g_{1,2}$ via Eqs.~(\ref{g12}). Thus, the solution of Eq.~(\ref{cq}), and the respective
form of $W(x)$, read:
%
\begin{align}
\rho(x)=\frac{{\rm tanh}(\alpha x)}{\sqrt{g_1{\rm tanh}^2(\alpha x)+g_2}}, ~~~
W(x)=\frac{2\epsilon\alpha g_1{\rm tanh}(\alpha x){\rm sech}^2(\alpha x)}{[g_1 {\rm tanh}^2(\alpha x)+g_2]^2},
\label{pteqn802}
\end{align}
where $\alpha^2=2\mu+\frac{1}{\epsilon^2}\pm\frac{1}{\epsilon}\sqrt{2\mu+\frac{1}{\epsilon^2}}$.
Next, let us consider the bright soliton of Eq.~(\ref{cq}),
$U(x)=\beta {\rm sech}(\alpha x)$; in this case, $l_0=g_2=0$ and, thus, we obtain
the solution $\rho(x)=1/\sqrt{g_1}$, which is
a constant function which does not produce a $\pt$-symmetric potential.
We finally note that other solutions of Eq.~(\ref{duf}) include periodic ones, in the form of
Jacobian elliptic functions, which can also be used to obtain $\pt$-symmetric potentials, but will not be considered here.

\section{Numerical Results}
In the previous section we presented analytical soliton solutions of Eq.~(\ref{pteqn1}),
as well as the respective $\pt$-symmetric potential $W(x)$, when the phase was taken to
be of the form $\phi_x=\epsilon\rho^k$ where $k=0,1,2$.
We now obtain numerically families of solutions for each $k$,
for different values of the parameter $\epsilon$ and then
study their linear stability.
This is done upon considering the perturbation ansatz
$u(x,t)=e^{-i \mu t} [u_0(x)+\delta (e^{\lambda t}{v}(x)+e^{\bar{\lambda} t}\bar{w}(x))]$,
where $\delta$ is a small parameter, and $(\lambda, \{{v},w\})$ denote eigenvalues and eigenfunctions (bar
denotes complex conjugate). Substituting this ansatz into Eq.~(\ref{pteqn1}), and linearizing with respect to $\delta$,
we derive
an eigenvalue problem, which is solved numerically.
%
Then, we study the evolution of unstable solutions with Re$(\lambda)>0$ by numerically integrating
Eq.~(\ref{pteqn1}), using a
4th-order Runge-Kutta scheme.
%
Notice that we do not consider the case of $k=0$, since it was shown to produce a non-physical form for the potential $W$.

\subsection{The cases $k=1$, $k=2$, and $\sigma=-1$ (focusing nonlinearity)}

\begin{figure}[tbp]
\centering
\begin{tabular}{cc}
\begin{tabular}{cc}
\hskip-0.6cm
\includegraphics[width=5cm]{\rootfig 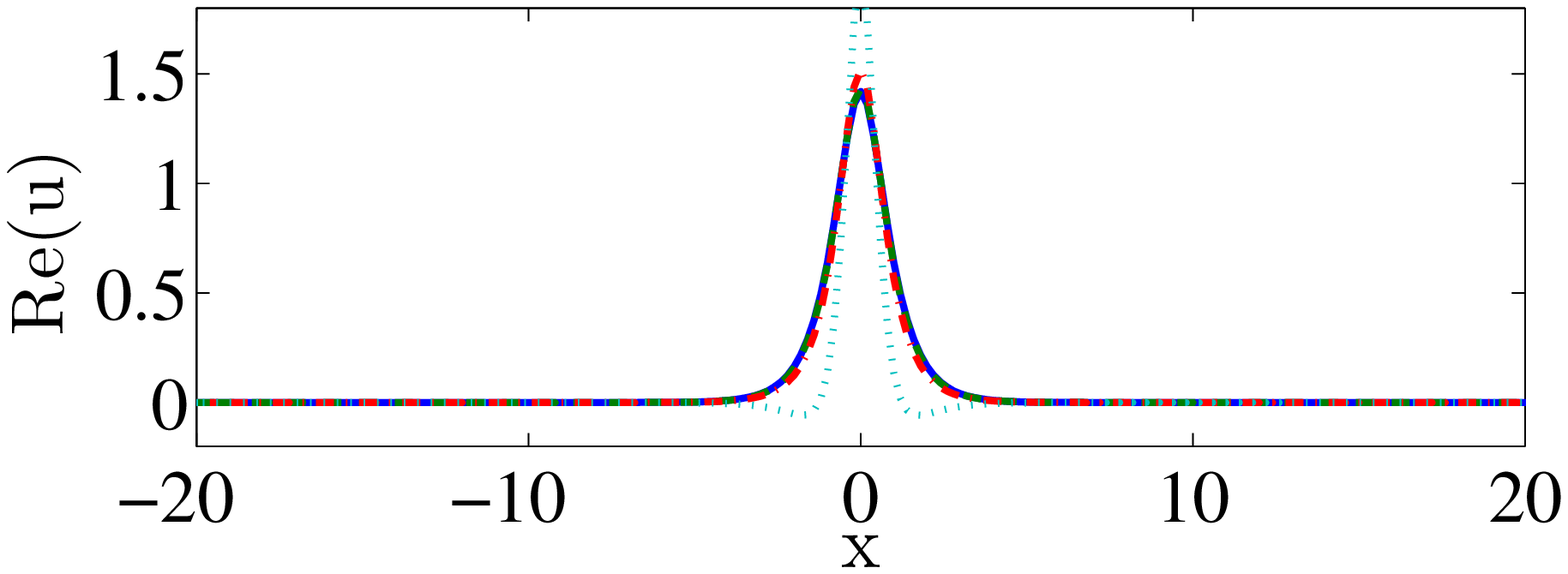}
&
\includegraphics[width=5cm]{\rootfig 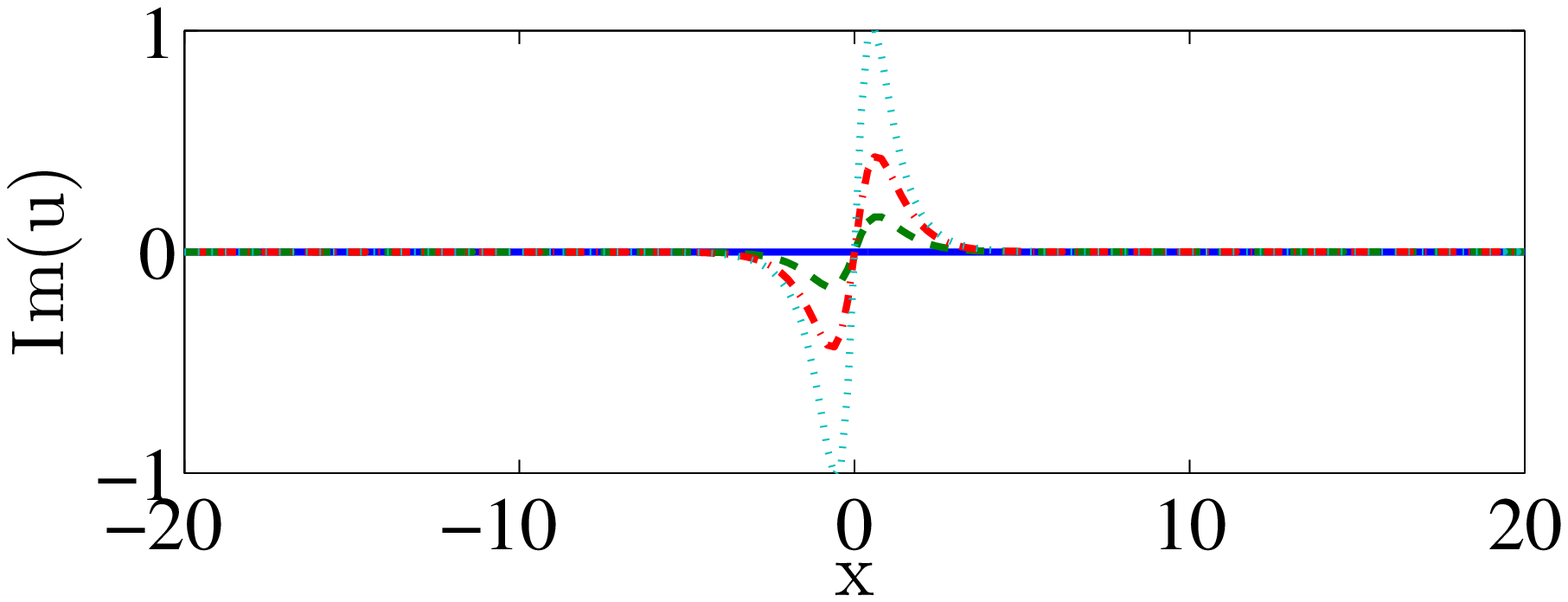}
\\
\hskip-0.6cm
\includegraphics[width=5cm]{\rootfig 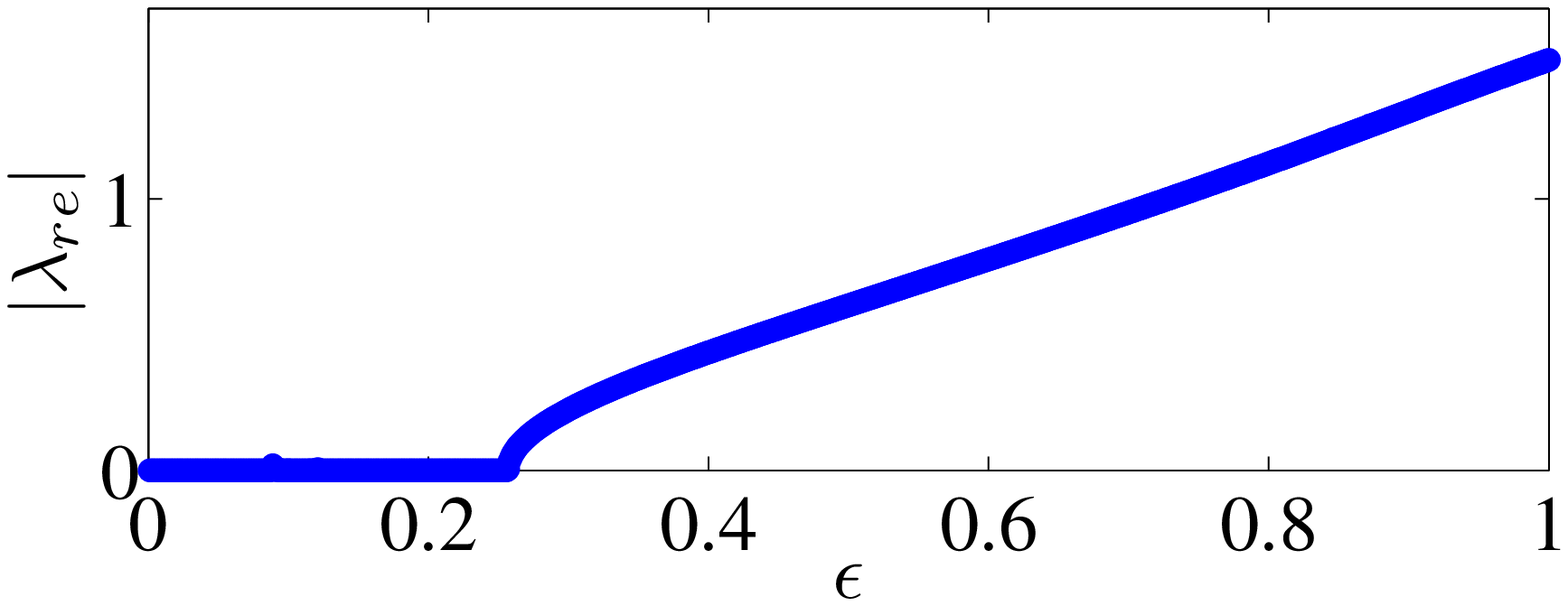}
&
\includegraphics[width=5cm]{\rootfig 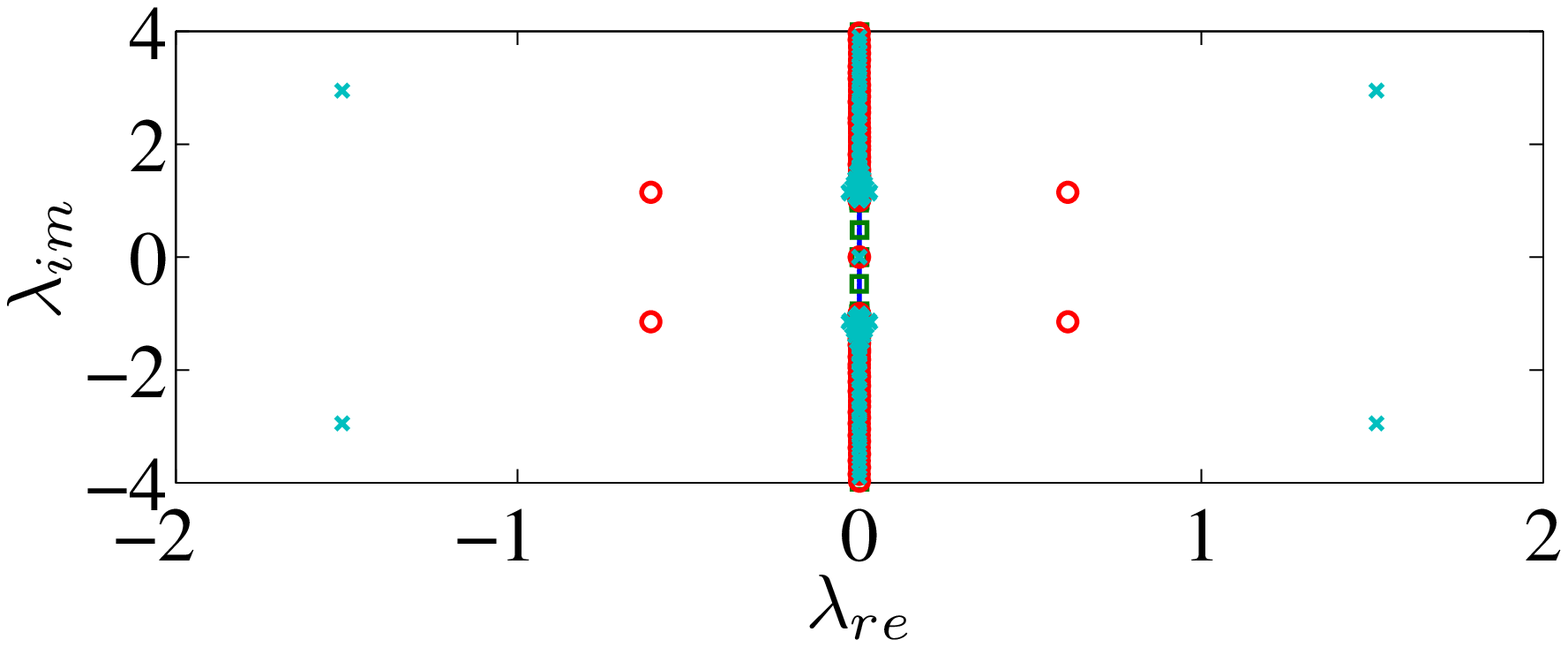}
\end{tabular}
\\[-4.4cm]
&
\hskip-0.30cm
\includegraphics[width=2.425cm]{\rootfig 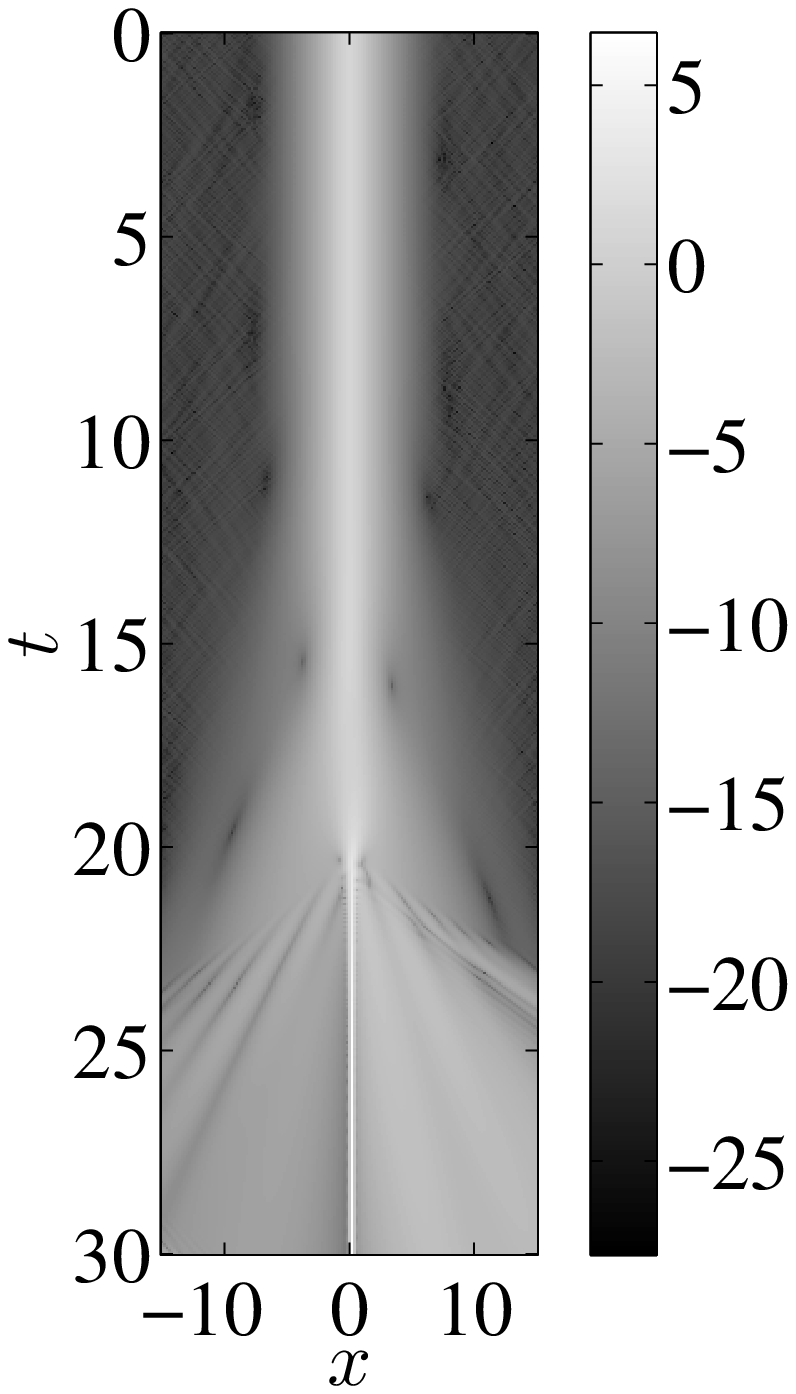}
\end{tabular}
\caption{Top
left and middle panels show, respectively, the profiles of Re$(u)$ and Im$(u)$ for
$\epsilon=0$ (solid), $\epsilon=0.2$ (dashed), $\epsilon=0.5$
(dashed-dotted), and $\epsilon=1$ (dotted).
The bottom left panel shows the maximal real part of the linearization
eigenvalues $\lambda$ as a function of $\epsilon$, showcasing the instability
beyond $\epsilon=0.257$. The bottom middle panel shows the spectral plane of
the imaginary vs. the real part of $\lambda$ for
$\epsilon=0$ (dots), $\epsilon=0.2$ (squares), $\epsilon=0.5$ (circles), and $\epsilon=1.0$ (crosses).
Right panel: Space-time contour plot of the evolution of the density $|u|^2$
(in logarithmic scale) of an unstable soliton for $\epsilon=0.4$.
}
\label{A}
\end{figure}

We start with the case $k=1$ and $\sigma=-1$, for which the density $\rho(x)$ has the form of
a bright soliton and the respective potential $W(x)$ is $\pt$-symmetric, as obtained in Eq.~(\ref{pteqn4}).
In the top left and middle
panels of Fig.~\ref{A}, we show the real and imaginary parts of the solution; it is observed that
the former remains symmetric, while its imaginary part develops an anti-symmetric profile
of increasing magnitude as $\epsilon$ increases.
On the other hand, see bottom left and middle panels
of Fig.~\ref{A}, it is interesting to observe the corresponding stability characteristics
of the solution. Rather unusually (especially for bright solitons in
single-component NLS
models), the spectrum features an {\it oscillatory} instability beyond the critical
value of $\epsilon=0.257$. The relevant instability eigenvalue
quartet features both a real and an imaginary part, indicating
the concurrent presence of growth (with a rate associated with the real
part) and oscillation (with a frequency associated with the imaginary part).
While such instabilities have been
identified previously, e.g., for bound states of two or more
solitons \cite{njpearly},
we are not aware of cases in Hamiltonian models, where the
breaking of translational invariance for a {\it single} soliton
due to the presence of an external potential leads to an oscillatory
instability. In the Hamiltonian case, this can be explained based
on the positive Krein signature (i.e., concavity of the energy
surface) along the eigendirection of the mode associated with translation
\cite{sanst}. Interestingly, the notion
of energy associated with a given mode or Krein signature
has not been extended
to the presence of gain and loss. Nevertheless, it can be clearly
seen that the eigenvalues still arise in pairs or quartets as
in the Hamiltonian case. This advocates both the relevance and
importance of developing an extension of the Krein signature
for such $\pt$-symmetric settings.

We now turn to direct numerical simulations exploring the dynamics
of such bright soliton solutions.
The right panel in Fig.~\ref{A} corresponds to
the unstable case of $\epsilon=0.4$.
The scale shown in the space-time contour plot is
logarithmic, so that the excessive growth associated with the
exponential instability does not obscure the development of
the oscillatory instability.


We note that we observed results similar to the ones presented above
for the case $k=2$ and $\sigma=-1$ (focusing nonlinearity); for instance,
%
when using the potential $W(x)=-4\sqrt{2}\epsilon\,\, {\rm sech}(\sqrt{2}x) \tanh(\sqrt{2}x)$, the only difference
we found lies in the location of the critical point
(occurring at $\epsilon=0.247$).

\subsection{The case $k=2$ and $\sigma=+1$ (defocusing nonlinearity)}

\begin{figure}[tb]
\centering
\begin{tabular}{cc}
\includegraphics[width=5cm]{\rootfig 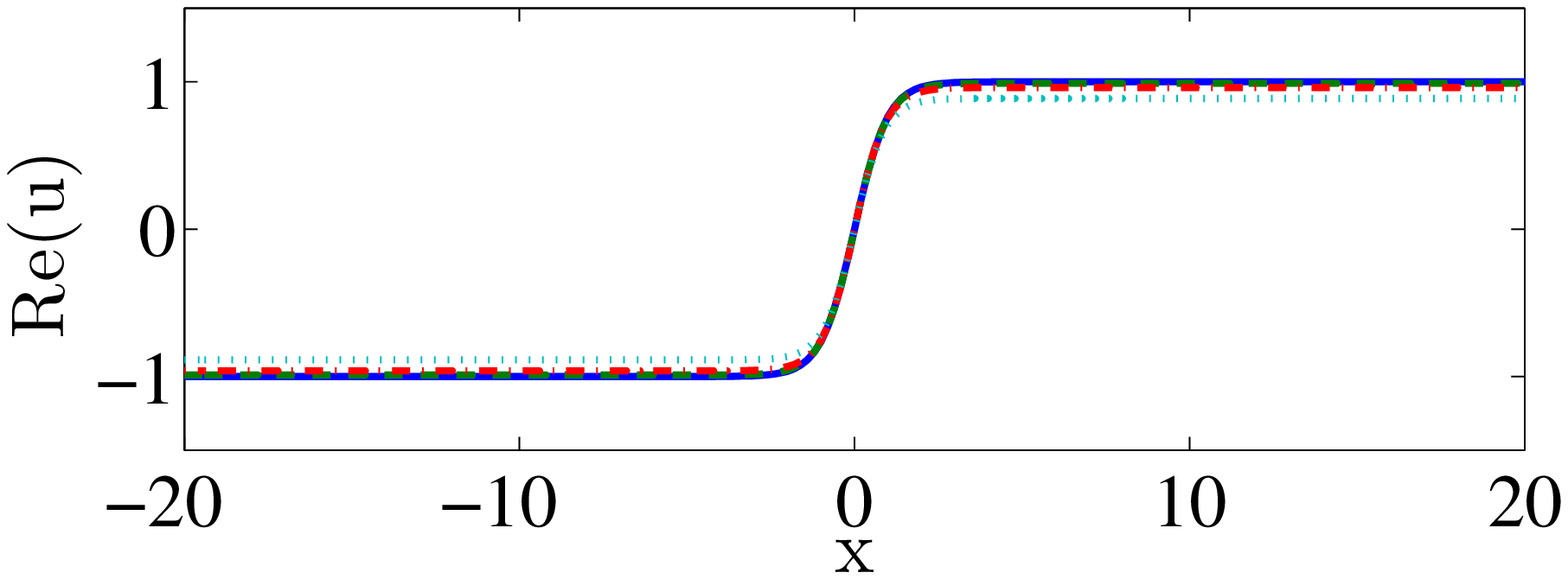}
\includegraphics[width=5cm]{\rootfig 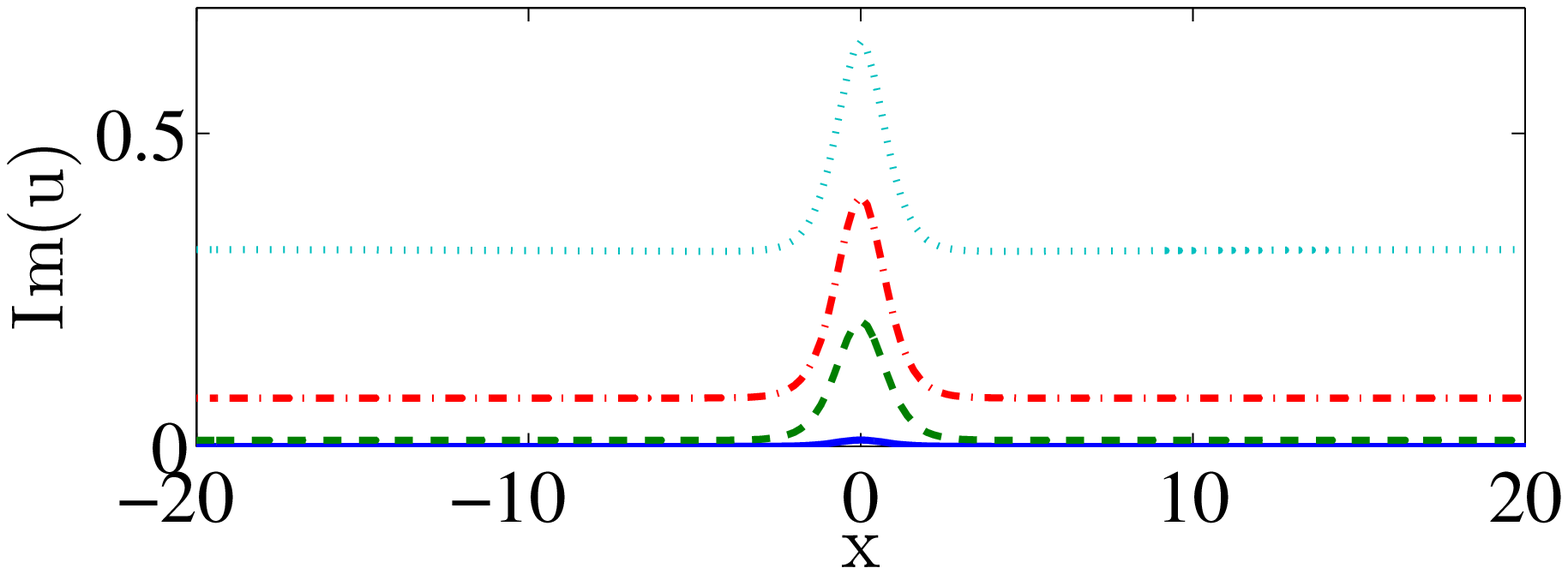}\\
\includegraphics[width=5cm]{\rootfig 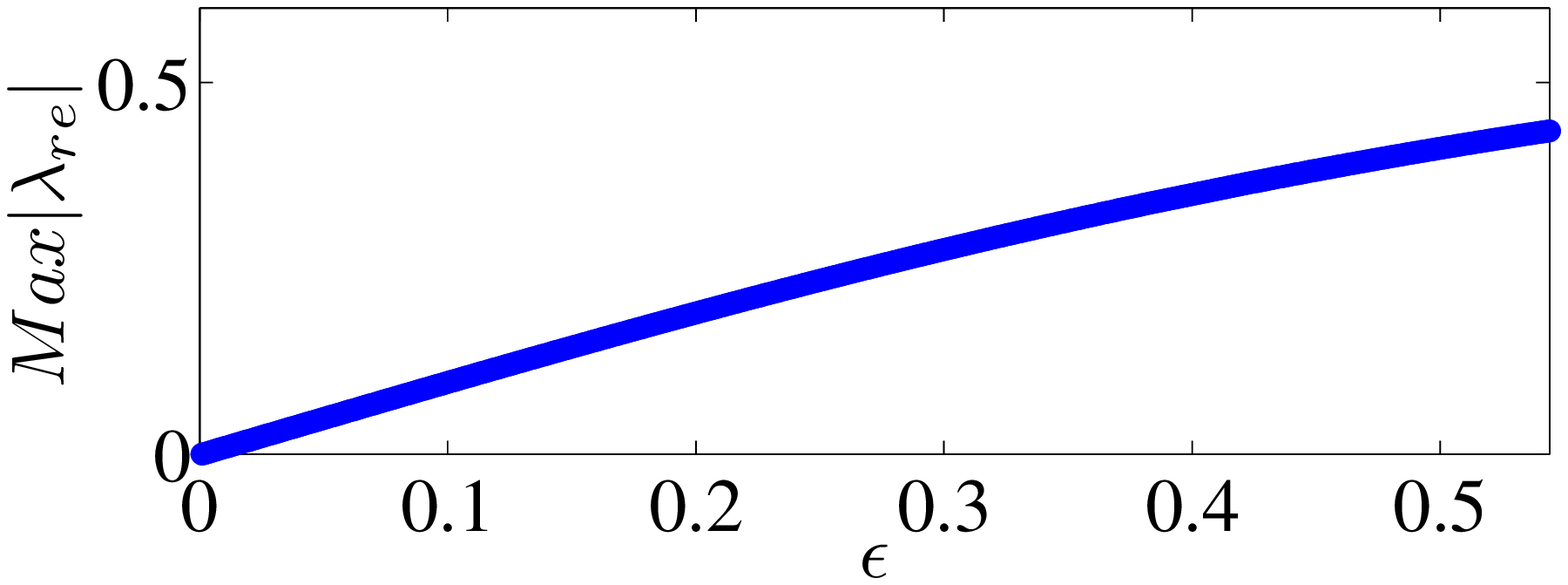}
\includegraphics[width=5cm]{\rootfig 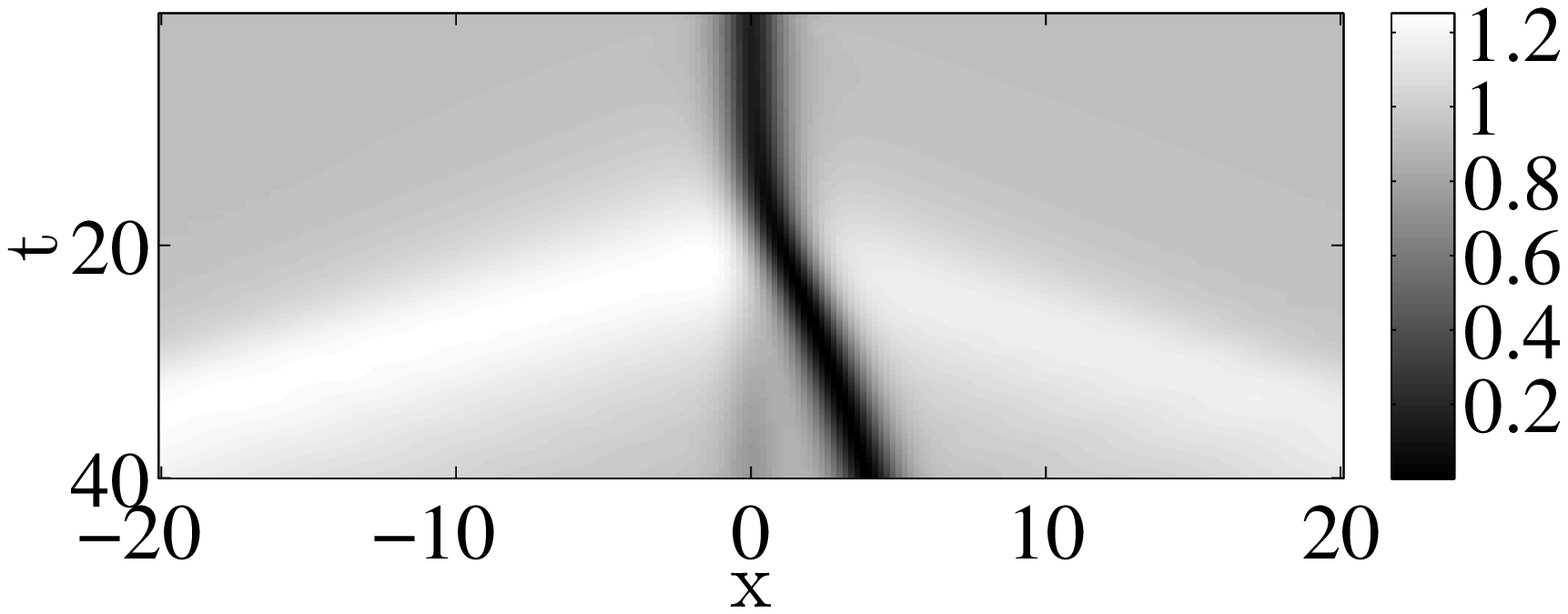}
\end{tabular}
\caption{
The top left (right) panel shows
the real (imaginary) part of the dark soliton solutions for $\epsilon=0.01$ (solid),
$\epsilon=0.2$ (dashed), $\epsilon=0.4$ (dashed-dotted), and $\epsilon=0.54$ (dotted).
The bottom
left panel shows the most unstable real eigenvalue as a function of
$\epsilon$.
The bottom right panel shows a space-time contour
plot of the evolution for $\epsilon=0.4$.}
\label{C_0}
\end{figure}

We now consider the case $k=2$ and $\sigma=+1$ (defocusing nonlinearity),
which
leads to the $\pt$-symmetric potential in Eq.~(\ref{pteqn802}).
The existence and stability results for the relevant dark soliton solutions are illustrated in Fig.~\ref{C_0}.
Here, similarly to the results of Ref.~\cite{Achilleos}, in the top
panels it is observed that
while the real part of the dark soliton remains anti-symmetric,
its imaginary part develops a symmetric, sech-shaped
profile, on the background of a homogeneous, non-vanishing pedestal.
As concerns their
stability, these states are
unstable due to the emergence of real eigenvalue pairs, for all values of $\epsilon$;
this family of solutions
can only be followed until $\epsilon=0.544$ (see below). The bottom left panel of Fig.~\ref{C_0}
shows that the maximum of the (absolute)
real part of the eigenvalues, is increasing with increasing $\epsilon$. Note that the imaginary part of the unstable eigenvalues was found to be zero for all $\epsilon$.
In particular, it is found that the
corresponding spectra (not shown here) exhibit increasingly more unstable eigenvalues, by increasing $\epsilon$.
The latter suggests that, not only the soliton but also the background becomes unstable.
An example of the instability dynamics is shown in the bottom right panel of the same figure, for $\epsilon=0.4$.
It is observed that the soliton remains quiescent for some time, but then it
is spontaneously ejected to the ``gain'' side of the imaginary potential (for $x>0$) and starts moving with
a finite velocity away from the origin.
This behavior can be qualitatively explained by the fact that the soliton
feels an effective potential \cite{Achilleos}.

\begin{figure}[tbp]
\centering
\includegraphics[width=4.25cm,height=1.8cm]{\rootfig 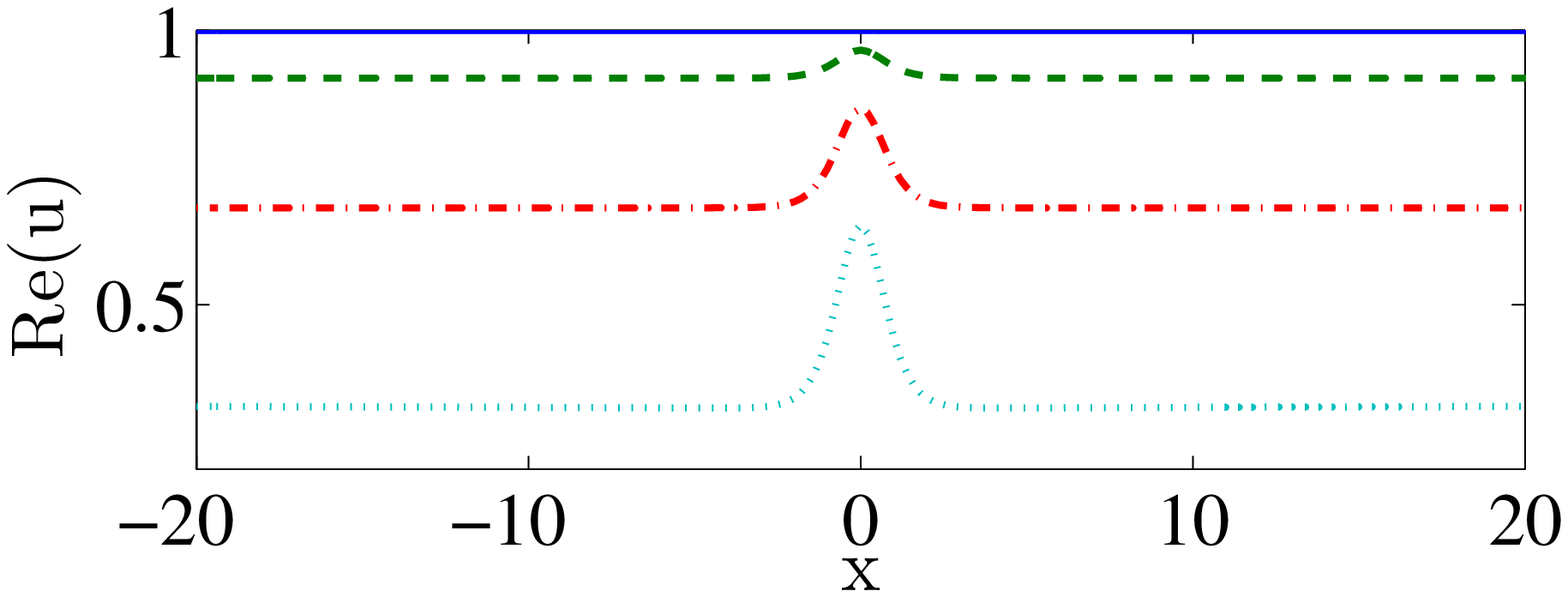}
\includegraphics[width=4.25cm,height=1.8cm]{\rootfig 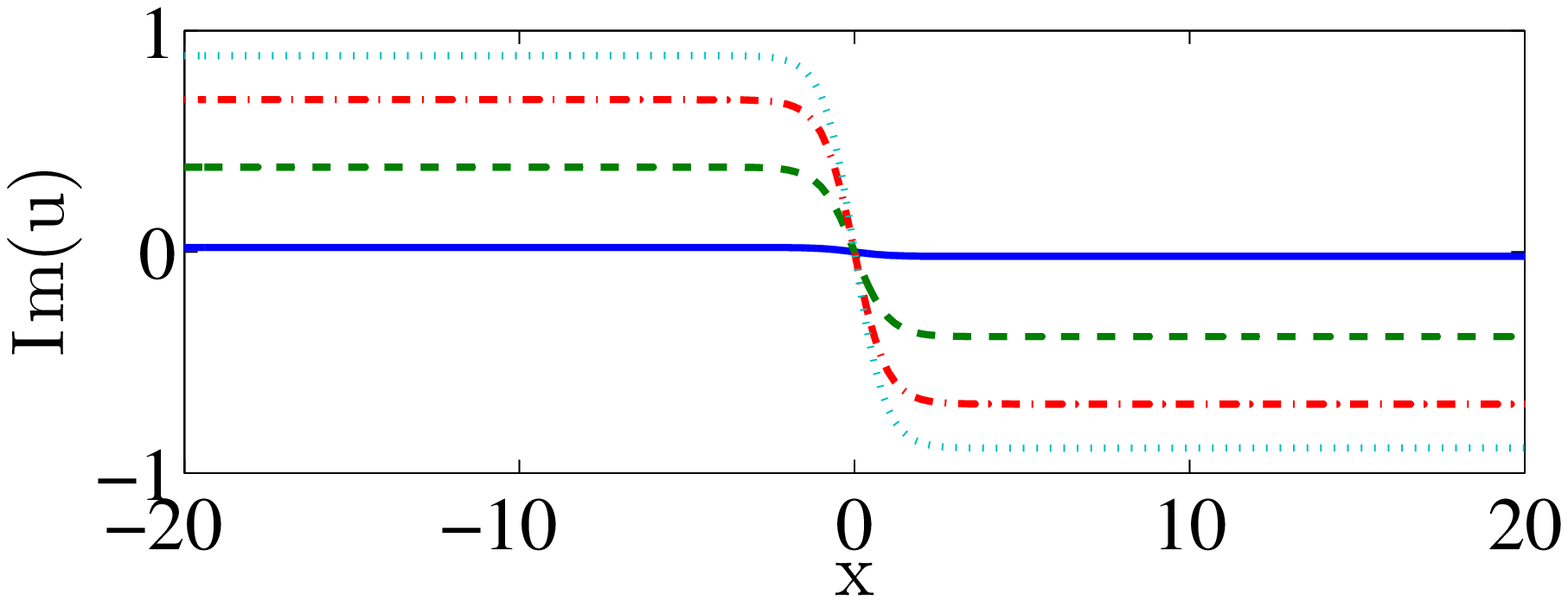}
\includegraphics[width=4.25cm,height=1.8cm]{\rootfig 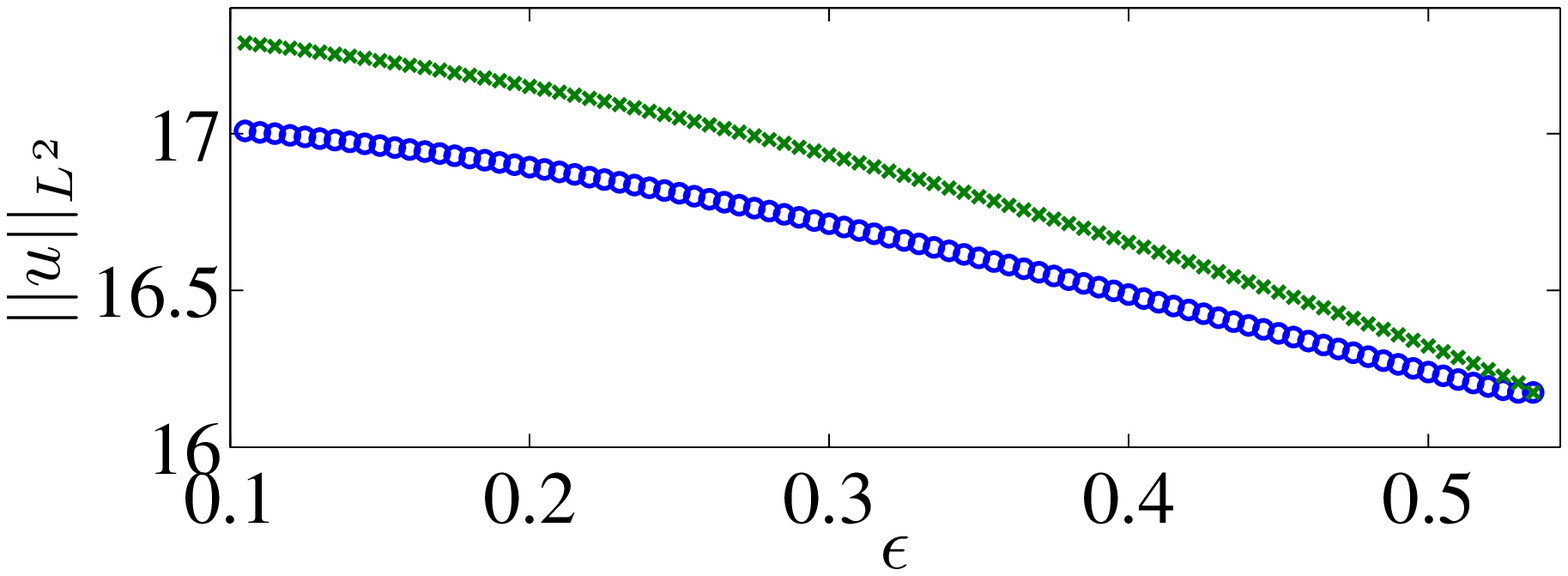}
\caption{
The left (middle)
panel shows
the real (imaginary) part of another branch of solutions for
$\epsilon=0$ (solid), $\epsilon=0.2$ (dashed), $\epsilon=0.4$ (dashed-dotted), and $\epsilon=0.54$ (dotted).
The right panel shows the bifurcation diagram of the $L^2$ norm of the branch
stemming from the homogeneous states
(green crosses) and the one
of the dark solitons
(blue circles);
the saddle-center bifurcation occurs at
$\epsilon=0.544$.}
\label{CD0}
\end{figure}

Additionally, as observed in the bottom right panel of Fig.~\ref{C_0}, at $x=0$ there exists a shallow, stationary localized dip for all times,
suggesting the existence of still another stationary, and in
this case stable, state. In fact, it is possible to find this branch,
for the same potential $W(x)$ [cf. Eq.~(\ref{pteqn802})], by performing numerical continuation in $\epsilon$,
starting from the plane wave solution of the NLS Eq.~(\ref{pteqn1})
---alternatively, this branch can be found in an approximate analytical form,
using the methodology of Ref.~\cite{Achilleos}.
This branch corresponds to the ground state of the system, and its real and imaginary parts
are respectively shown in the left and middle panels of Fig.~\ref{CD0}. Note that the real part of the solutions
has a sech-shaped bump localized at the origin (similar to the imaginary part of the dark soliton), while
its imaginary part is tanh-shaped (similar to the real part of the dark soliton).
As is expected, the ground state branch coexists with the excited sate (the soliton),
up to a certain critical value of $\epsilon$; at this value, the branches
collide and disappear via a saddle-center bifurcation, in a way similar to
the nonlinear $\pt$-phase transition \cite{Achilleos}. This is illustrated in the right
panel of Fig.~\ref{CD0}, where the norms $\|u\|_{L^2}$ of the above mentioned branches are shown
as functions of $\epsilon$ (the critical value of $\epsilon$ is found to be $\epsilon=0.544$).
%

\subsection{The case of a modified $\pt$-symmetric potential}

\begin{figure}[tbp]
\centering
\includegraphics[width=4.3cm,height=1.8cm]{\rootfig 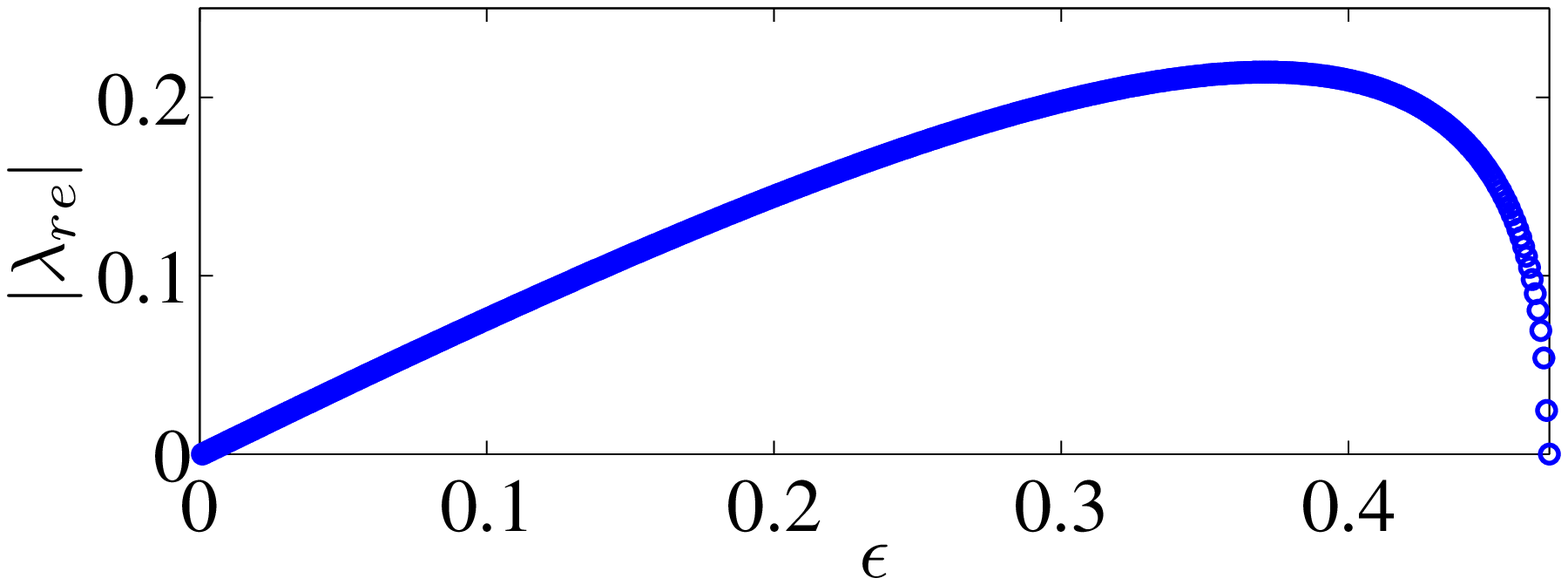}
\includegraphics[width=4.2cm,height=1.9cm]{\rootfig 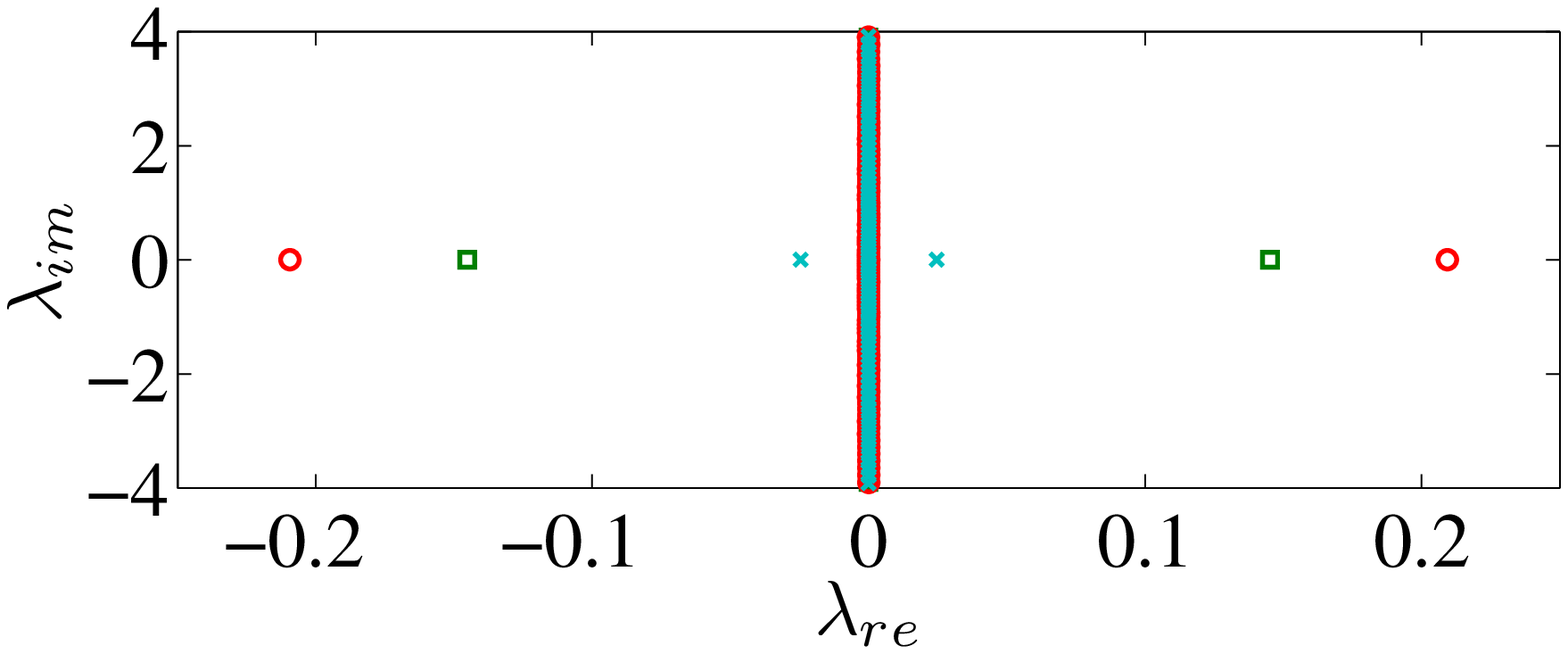}
\includegraphics[width=4.2cm,height=1.85cm]{\rootfig 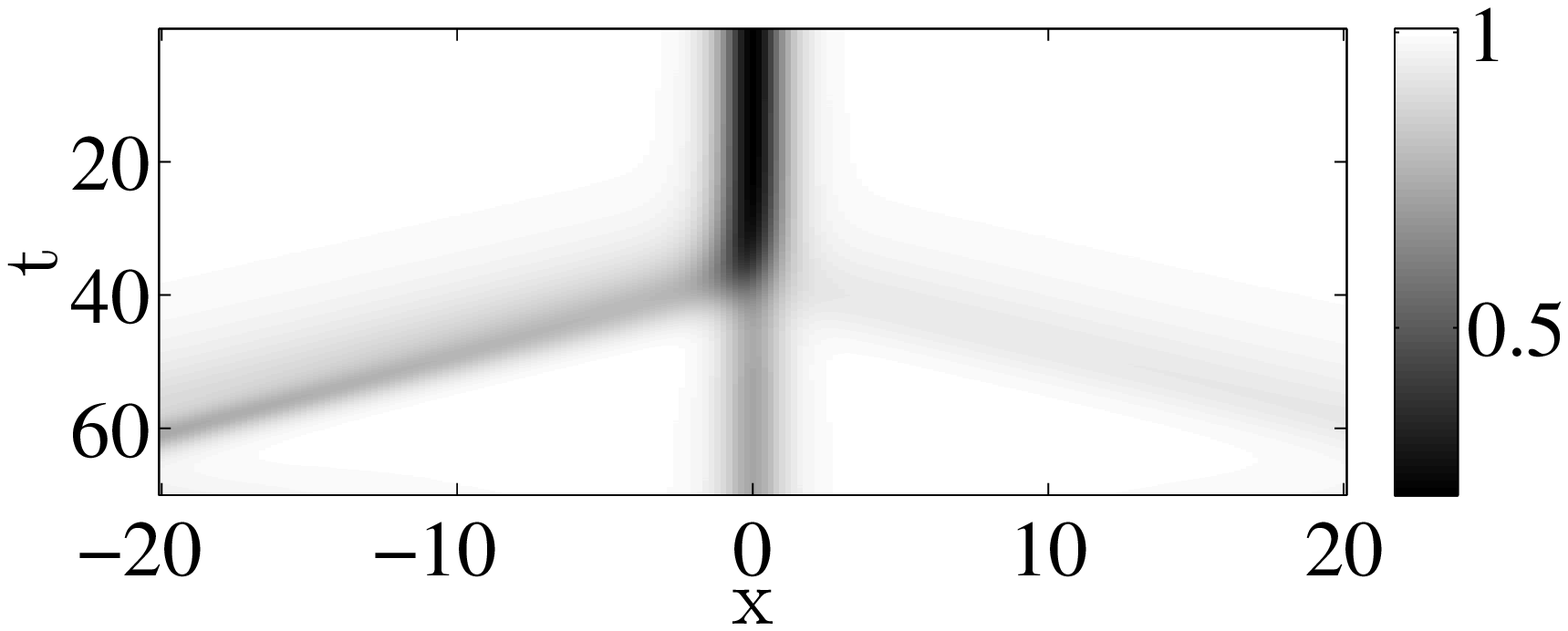}
\caption{
Left panel: the most unstable real eigenvalue,
performing a maximal excursion over
$\epsilon$, before returning to collide with the
origin at $\epsilon=0.469$. 
Middle panel:
spectral plane for $\epsilon=0$ (dots), $0.2$ (squares), $0.4$ (circles),
 $0.469$ (crosses).
Right panel: space-time contour
plot showing
the evolution of the soliton density
for $\epsilon=0.4$.
 }
\label{C}
\end{figure}

In the previous case of $k=2$, and for $W(x)$ given by Eq.~(\ref{pteqn802}), we found that soliton solutions
gradually became highly unstable due to the fact that, not only the solitons, but also their backgrounds became unstable.
For this reason, we investigate a modified version of $W(x)$, namely the
$\pt$-symmetric potential $W(x)=2\epsilon\,\, {\rm sech}^2(x)\tanh(x)$. Although exact analytical dark soliton solutions
are not available in this setting, one can obtain soliton solutions numerically. In fact, we have found such a soliton
branch, with profiles similar to those shown in the top panels of Fig.~\ref{C_0}; we also found that
this branch disappears at
$\epsilon=0.469$. Accordingly, a stability analysis for this soliton branch shows (see left panel of Fig.~\ref{C}) that the maximum unstable
(real) eigenvalue departs from the origin, performs a maximal excursion and subsequently return to the origin
at $\epsilon=0.469$,
in a way reminiscent of a bifurcation. An additional
important difference
of
these soliton states with the ones corresponding to $W(x)$ in Eq.~(\ref{pteqn802}) is that, now,
the instability is caused by a {\it single} real eigenvalue pair ---cf. middle panel of Fig.~\ref{C}.
This eigenvalue pair, is associated with the motion of the dark soliton \cite{Achilleos}; the absence of
other unstable eigenvalues suggests that, while the dark soliton is unstable, the background is not
---unlike the previous case of $W(x)$ in Eq.~(\ref{pteqn802}), where the presence of many unstable modes
indicated the instability of the background.
%

An
example of the dynamics of an unstable soliton evolving in the presence of
the modified potential $W(x)$, is shown in the right panel of
Fig.~\ref{C}.
%
Evidently, the instability renders the
soliton mobile,
in a similar way as in the bottom right panel of Fig.~\ref{C_0}.
Additionally, as in the case of $W(x)$ given by Eq.~(\ref{pteqn802}),
there exists a persistent localized structure at $x=0$, adjacent to the soliton.
As before, we can associate this structure with the existence of another
branch of solutions, which can be obtained by a continuation, starting from a plane
wave solution at $\epsilon=0$.
We have numerically obtained
this ground state branch, and it was found to have a similar profile as the one shown in
the left and middle panels of Fig.~\ref{CD0}. We have also confirmed that this
branch collides with the soliton branch at $\epsilon=0.469$, where they disappear
through a saddle-center bifurcation, similar to what is shown in the right panel
of Fig.~\ref{CD0}. It is important to note that, contrary to the previous case
of $W(x)$ given by Eq.~(\ref{pteqn802}), this branch is found to be {\it stable} throughout
its existence regime; furthermore, we have confirmed that the localized structure
at $x=0$,
shown in the right panel of Fig.~\ref{C}, is {\it identical} to the exact solution
profile of the ground state branch.

\section{Conclusions and Future Challenges}

Concluding, we have studied a class of
$\pt$-symmetric NLS models, in which
their hydrodynamic form
can be reduced to a
Duffing or a generalized Duffing equation. From this type of reduction, we
were able to extract a number of cases bearing exact analytical soliton
solutions. The stability and dynamics of these solutions were studied numerically,
and a number of intriguing features were found.
In particular, we presented
cases where bright solitons in a single-component, $\pt$-symmetric
NLS equation are subject to an oscillatory instability, a trait absent in the
Hamiltonian installments of the model. This underscored the
need for generalizing the notion of Krein signature in such
systems. On the other hand, for defocusing nonlinearities,
a saddle-center bifurcation reminiscent of
the $\pt$-phase transition was found between the ground
state
(plane wave in the Hamiltonian limit) and the first
excited state (a dark soliton in the Hamiltonian limit).
Connections of all these features to earlier works were also
given. We also note that we have studied
cnoidal wave solutions (results not shown), which
were found to be dynamically unstable.

There are many interesting future directions associated with this work.
In the 1D setting, it is important
to understand
topological notions,
such as the Krein signature, and other stability characteristics
of solutions of $\pt$-symmetric systems, as well as their connections to the Hamiltonian analogs.
%
On the other hand, admittedly even for the existence problem, there
are some intriguing questions (such as the difference in the
bifurcation structure between the cases of focusing and defocusing nonlinearities).
Furthermore, while numerous studies have, by now, developed
in the 1D
realm, two- (and higher-)
dimensional explorations are presently far more limited.
Hence, this is also a direction of considerable interest for future
studies.

\end{document}